\documentclass[]{JHEP}
\usepackage{epsfig}
\def\Tr{\mathop{\mbox{Tr}}\,}

\def\di{\mbox{d}}
\newcommand{\bea}{\begin{eqnarray}}
\newcommand{\be}{\begin{equation}}
\newcommand{\eea}{\end{eqnarray}}
\newcommand{\ee}{\end{equation}}
\newcommand{\nn}{\nonumber}

\title{Supergravity and matrix theory do not disagree 
on multi-graviton scattering}

\author{Marco Fabbrichesi, Gabriele Ferretti and Roberto Iengo\\
INFN, Sezione di Trieste and\\
Scuola Internazionale Superiore di Studi Avanzati (SISSA)\\
via Beirut 4, I-34013 Trieste, Italy.}

\abstract{We compare the  
amplitudes for the long-distance scattering of three
gravitons in eleven dimensional 
supergravity and  matrix theory at finite $N$.
We show that the leading supergravity term arises from loop contributions to
the matrix theory effective action that are not required to vanish by
supersymmetry.
We evaluate in detail one type of diagram---the setting sun with only
massive propagators---reproducing the supergravity behavior.}


\preprint{SISSA 39/98/EP \\
April 1998}
\begin{document}
\section{Introduction}                    

In the original work of BFSS~\cite{BFSS}, it was conjectured
that M theory could be
described by a matrix model in the large $N$ limit; later,
refs.~\cite{Su,sen,sei} gave meaning to the matrix model also at finite $N$ as
describing M theory on the discrete light-cone. Even though,
in the long-distance regime, M theory is supergravity in eleven dimensions, 
there is no guarantee that the supergravity result will always match 
matrix theory at finite $N$ due 
to the presence of a small scale in the problem~\cite{PH}. 
For this reason, it is of importance to test in actual 
computations how far the two agree.   

The most convincing
test to date has been the comparison of the two-body 
scattering~\cite{BB,BBPT}.~\footnote{ For earlier works 
on different aspects of this problem, see~\cite{B,DKPS,SQM}.}
An even more stringent test would come from
multi-particle scattering. 
The case of three  11-dimensional 
gravitons, carrying Kaluza-Klein momentum in the $10^{\rm th}$
compactified direction, has been considered
in~\cite{DR} where the authors claim that a term present in the supergravity
amplitude cannot arise in the matrix model. 

In this note we discuss this important issue by reconsidering the computation
of~\cite{DR}. Our result is encouraging for matrix theory: contrarily to
what reported, we find that there are (0+1)-dimensional 
Yang-Mills (YM) graphs which 
lead to the same behavior with respect to the relative distances and
relative velocities as in the supergravity result. 

The diagrams we have considered
are two-loop  diagrams in the bosonic sector---there are
various similar diagrams which can give rise to the same 
behavior---and we have analyzed in detail one of them, the {\it setting-sun}
diagram with all massive propagators,
which only arises in the three-body problem. 

When considered in the framework of the effective action 
arising at one loop
by integrating out the long-distance degrees of freedom
(the {\it heavy} modes), our result originates from a term
that does not vanish after summing over all bosonic and fermionic
contributions---even though the effective operator for the
remaining {\it light} modes does not contain any explicit dependence on
the velocity. 
In the final two-loop effective action, obtained after integrating out all
modes, the velocity independent terms cancel (in agreement with 
supersymmetry~\cite{D2,PSS})
while, the relevant contribution arises  from a term of
order $v^6$ which is not expected to vanish by supersymmetry. 

We have not attempted to compute the numerical coefficient, 
which would require the
algebraic sum of the various bosonic, fermionic and ghost diagrams of
the YM theory at order $v^6$. 
Therefore we present here a {\it minimal} result, which we feel nonetheless to 
be important because of the recent discussion 
concerning whether one could or could not find  the
supergravity behavior in the  diagrams of the YM formulation
of matrix theory at finite $N$. 

\section{The amplitude in supergravity}   

The simplest way to obtain supergravity amplitudes is by means of string
theory. Since it is a tree-level amplitude, it is consistent with
conformal invariance in any
dimensionality, in particular in $D=11$. We consider the {\it bona fide} 
superstring theory (where there is no tachyon) and the scattering amplitude
of three ($11$-dimensional) gravitons, and look at suitable {\it pinching}
limits,
where only intermediate massless states are coupled to the external
gravitons. Those states are themselves $11$-dimensional gravitons.
We then compactify the $10^{\rm th}$ space dimension giving mass
to the external gravitons, which will thus correspond to
$10$-dimensional $D0$-branes. Keeping zero momentum transfer in
the  $10^{\rm th}$ dimension, the intermediate states remain massless
and correspond to the various massless fields of $10$-dimensional
supergravity.
 
The supergravity amplitude is thus obtained from that of 
superstring theory
by a limiting procedure that isolates the relevant corners of the moduli
space. We follow \cite{FIR}, where the appropriate technology is explicitly 
developed. 

By  considering only the part of the complete amplitude that is
proportional to 
\be
\varepsilon_1 \cdot \varepsilon_1' \: 
\varepsilon_2 \cdot \varepsilon_2' \: \varepsilon_3 \cdot \varepsilon_3' \, ,
\ee
$\varepsilon$ being the external graviton polarization tensor,
we obtain the amplitude $A_6$ for six graviton vertices:
\bea
A_6 & = & \varepsilon_1 \cdot \varepsilon_1' \: 
\varepsilon_2 \cdot \varepsilon_2' \: \varepsilon_3 \cdot \varepsilon_3' \:
\frac{\kappa^4 (\alpha')^3}{4 \pi^3} \int \di ^2 x\: \di ^2 y\: \di z^2 
 |1-y|^{-2 + \alpha' p_2'\cdot p_2} \nn \\
&&\times \: |y|^{\alpha' p_3\cdot p_2'} 
|1-x|^{\alpha' p_2\cdot p_1'} |x|^{\alpha' p_3\cdot p_1'}
 |1-z|^{\alpha' p_3'\cdot p_2} \nn \\
&&\times \:  |z|^{-2 + \alpha' p_3\cdot p_3'}
|z-x|^{\alpha' p_3'\cdot p_1'} |z-y|^{\alpha' p_3'\cdot p_2'}
 |x-y|^{\alpha' p_2'\cdot p_1'} \nn \\
&&\times \: \left\{ \frac{p_3' \cdot p_1' \: p_2' \cdot p_1'}{(y-x)(z-x)} +
\frac{p_3 \cdot p_2' \: p_3' \cdot p_1'}{y(z-x)} -
\frac{p_3' \cdot p_2' \: p_3 \cdot p_1'}{x(z-y)} \right. \nn \\
&& \left. + \frac{p_2' \cdot p_3' \: p_2' \cdot p_1'}{(y-x)(z-y)} +
\frac{p_3' \cdot p_2 \: p_2' \cdot p_1'}{(z-1)(y-z)} \right\} 
\wedge \Biggl\{ c.c. 
\Biggr\} \, . \label{a6-string}
\eea
The eleven-dimensional momenta are chosen to be
\be
p_i = (E_i, {\bf p}_i-{\bf q}_i /2, M_i) \quad \quad 
p_i' = (-E_i', - {\bf p}_i-{\bf q}_i /2, -M_i) \label{momenta}
\ee
where $p_i^2=0$, $E_i \simeq M_i + ({\bf p}_i-{\bf q}_i /2)^2/2M_i$ and
$M_i=N_i/R_{11}$
are the momenta in the compactified dimension. Energy-momentum
conservation gives $\sum_i {\bf q}_i = 0$ and
$\sum_i {\bf p}_i \cdot {\bf q}_i = 0$.

In order to obtain a non-vanishing result in the field theory limit ($\alpha'
\rightarrow 0$), we must extract three poles in momenta, each of them 
bringing down one power of  $(\alpha')^{-1}$ and thus compensating for
the three powers of $\alpha'$ in front of $A_6$; each pole 
originates from a pinching limit in which some of the Koba-Nielsen variables
come close to each other. 
The pinching limits corresponding to the grouping of the six external
vertices into three pairs $i,i'$,  with $i=1$, $2$ and 3, give field theory
diagrams where the incoming ($i$) and outgoing ($i'$) particles of each pair
describe the world-line of the $D0$-brane number $i$.

In particular, we are interested in seven pinching limits. 
One of them corresponds to the so-called Y diagram where each of the three
world-lines are coupled to one intermediate massless state, and the three
intermediate states meet at a point. We call the
corresponding amplitude $A_{Y}$. In addition, there are the diagrams
where one world-line is coupled to two intermediate states, each of
them attached to one of the two other world-lines (with six possible choices).
We disregard terms, which are interpreted as {\it re-scattering} effects,
where a world-line interacts successively with two intermediate states,
with an external particle's propagator in between. Thus,  
besides the Y diagram,  we are left
with diagrams where two intermediate states originate from the same point
of one of the world-lines.  We denote the 
corresponding amplitude by $A_{\vee}$ and, therefore, we have that
$A_6 = A_Y + A_\vee$. We keep 
only those terms giving the maximal singularity in the momentum transfers.

Let us first consider the amplitude (taking, for the moment, $N_i=1$) 
\bea
A_\vee & = & 2 \: \kappa^4 \: \varepsilon_1 \cdot \varepsilon_1' \: 
\varepsilon_2 \cdot \varepsilon_2' \: \varepsilon_3 \cdot \varepsilon_3' \nn \\
&& \times \left\{
 \frac{({\bf p}_3 - {\bf p}_2)^2 \: ({\bf p}_3 - {\bf p}_1)^2
\: ({\bf p}_2 - {\bf p}_1)^2}
{{\bf q}_1^2\: {\bf q}_2^2\: {\bf q}_3^2}  \left( \frac{{\bf q}_1^2 + 
{\bf q}_2^2 + {\bf q}_3^2}{2}
\right)  \right\} \, . \label{a6-sugra}
\eea
Eq.~(\ref{a6-sugra})
is the same supergravity amplitude considered in~\cite{DR}. 

Eq.~(\ref{a6-sugra}) contains three possible 
singular configurations in which two of the ${\bf q}_i$'s are small,
describing by Fourier transform the large-distance interaction
of one $D0$-brane with the other two.  Consider the case where the
distance of the brane number $1$ from the branes
number $2$ or $3$ is much larger than the distance between
the branes $2$ and $3$. The two singular terms proportional to 
\be
\frac{1}{{\bf q}_1^2 {\bf q}_2^2} \quad \mbox{or} \quad 
\frac{1}{{\bf q}_1^2 {\bf q}_3^2}   \label{onefar}
\ee 
are relevant. Let us consider the first one, as the second one is
obtained by interchanging $2$ with $3$. The Fourier transform 
gives (${\bf r}_i$ being the  position in space of the $i$-th brane)
\bea
a_\vee &=&  2 \: \kappa^4  \: \varepsilon_1 \cdot \varepsilon_1' \: 
\varepsilon_2 \cdot \varepsilon_2' \: \varepsilon_3 \cdot \varepsilon_3' 
\: \: ({\bf p}_3 - {\bf p}_2)^2 \: ({\bf p}_3 - {\bf p}_1)^2
\: ({\bf p}_2 - {\bf p}_1)^2 \nn \\
& & \times \int \frac{\di^9{\bf q}_1 \di^9{\bf q}_2}{(2\pi)^{18}} 
\frac{1}{{\bf q}_1^2\: {\bf q}_2^2} 
\: \exp \Bigl[ i \: {\bf q}_1 \cdot  ({\bf r}_1 - {\bf r}_3) 
+   i \: {\bf q}_2 \cdot  ({\bf r}_2 - {\bf r}_3)\Bigr] 
\eea

To get the case of  generic $N_i$, we have to replace 
\be
({\bf p}_i - {\bf p}_j)^2 \rightarrow 
\frac{M_j}{M_i}{\bf p}_i^2 +\frac{M_i}{M_j}{\bf p}_j^2 
-2{\bf p}_i\cdot {\bf p}_j
\ee
and write the momenta in terms of the 
velocities as ${\bf p}_i =M_i {\bf v}_i$ while
bearing in mind that $M_i\sim N_i$. 
 We normalize the amplitude by dividing the result
by the product of the $M_i$
and obtain:
\be
a_\vee \sim   
\frac{N_1 N_2 N_3 ({\bf v}_3 -  {\bf v}_2)^2 ({\bf v}_3 - {\bf v}_1)^2 
({\bf v}_2 - {\bf v}_1)^2} 
{|{\bf r}_1 -{\bf r}_2|^7 |{\bf r}_2- {\bf r}_3|^7} \label{a6} \, .
\ee 

In order to compare~(\ref{a6})
with matrix theory we consider the {\it eikonal expression}
where we integrate over the time $t$ along the world-line trajectories. 
For simplicity we take the velocities of all three particles to be along 
the $X^1$ axis and the relative displacements to be purely transverse 
(impact parameters). In other words, for the $i$-th particle,
${\bf r}_{i} = (v_i {\bf\hat n}_1 t +{\bf b}_{i})$, with 
${\bf b}_i\cdot{\bf\hat n}_1=0 $.
This integral gives, in the limit where 
$B\equiv |{\bf b}_1 -{\bf b}_2| \gg b \equiv |{\bf b}_2 -{\bf b}_3|$,
\be
\tilde{a}_\vee  \sim \int \di t\; 
\frac{N_1 N_2 N_3 v_{23}^2 v_{13}^2 v_{12}^2}{(v_{23}^2t^2 + B^2)^{7/2}
(v_{12}^2t^2 + b^2)^{7/2}}\sim 
\frac{N_1 N_2 N_3 |v_{23}| v_{13}^2 v_{12}^2} 
{B^7 b^6} \, , \label{sugra}
\ee
where $v_{ij} \equiv v_i - v_j$.
The other term in  (\ref{onefar}) gives the same amplitude with $B$  
replaced by  
$B'\equiv |{\bf b}_1 -{\bf b}_3| \simeq B$. It is the amplitude
$\tilde{a}_\vee$ in~(\ref{sugra})
 that we want to reproduce in the matrix theory computation.

As for the Y diagram, the corresponding eikonal expression $\tilde{a}_Y$
turns out to be sub-leading in our limit (see the appendix A).
 
\section{The amplitude in matrix theory} 

The derivation of the Feynman rules and the computation of the relevant 
diagrams follow closely those of \cite{BB}. We use  units where
\be
g_{\mbox{\rm \scriptsize YM}}=\left( R_{11}/
\lambda_{\mbox{\rm \scriptsize P}}^2 \right) ^{3/2}=1 \, ,
\ee
the quantities $R_{11}$, $\lambda_{\mbox{\rm \scriptsize P}}$ 
and $g_{\mbox{\rm \scriptsize YM}}$ 
being the 
compactification radius, the Planck length and the Yang-Mills
coupling, respectively. The bosonic part of the gauge fixed action reads 
\cite{BB} 
\bea
S &=& \int \di t \: \:\Tr\bigg(\dot a_0^2 + \dot x_i^2 + 
4\,i\,\dot R_k\,[a_0, x_k] 
-[R_k, a_0]^2 - [R_k, x_j]^2\nn\\
&&+2\,i\,\dot x_k\,[a_0, x_k] + 2\,[R_k, a_0][a_0, x_k] 
-2\,[R_k, x_j][x_k, x_j] \nn\\
&&-[a_0,x_k]^2 - \frac{1}{2}[x_k, x_j]^2 \bigg), \label{action}
\eea
where $a_0$ and $x_k$ are hermitian matrices representing the fluctuations 
and $R_k$ is the background.
Since we are studying the scattering of three $D0$-branes, 
with two independent velocities and impact parameters,
we need to consider at least 
a rank two group, namely $SU(3)$. The overall factors of
$N_{i}$, representing the longitudinal momentum will be fixed at the end.

We choose the same background that led to (\ref{sugra}), namely,
\be
   R_1 =\pmatrix{v_1 t & 0     & 0     \cr
                 0     & v_2 t & 0     \cr
                 0     & 0     & v_3 t \cr} \qquad\hbox{and}\qquad
   R_k =\pmatrix{b_k^1   & 0       & 0       \cr
                 0       & b_k^2   & 0       \cr
                 0       & 0       & b_k^3   \cr}\quad k>1.
\ee

We factor out the motion 
of the center of mass by imposing $v_1 + v_2 + v_3 = 0$ and
${\bf b}^1 +{\bf b}^2 +{\bf b}^3 = 0$. 

We use a Cartan basis for $SU(3)$, where $H^1$ and $H^2$ denote the 
generators of the Cartan sub-algebra and $E_\alpha$ ($\alpha=\pm\alpha^1,
\pm\alpha^2,\pm\alpha^3$) the roots. We also define the space vectors
\be
   {\bf R}^\alpha = \sum_{a=1,2}\alpha^a\Tr \Big(H^a {\bf R}\Big).
\ee
With the standard choice of $H^a$ and $\alpha$, this definition singles out 
the relative velocities and impact parameters, e.g. 
$ R_1^{\alpha^1} = (v_2 - v_3)t\equiv v^{23}t$ 
and, for $k>1$, $ R_k^{\alpha^1} = b_k^2 - b_k^3\equiv b_k^{23}$ 
together with cyclic permutations. 
According to the previous section we choose the 
relative distance of the first particle with the other two to be much larger 
than the relative distance of particle two and three, in other words, we set
\be
   |{\bf b}^{\alpha^2}|\approx|{\bf b}^{\alpha^3}|\approx B \gg
   |{\bf b}^{\alpha^1}|\approx b \quad \mbox{and} \quad 
B\, b \gg v \, , \label{regime}
\ee 
where in our units $v$ has the same dimensions as $b^2$.

The propagators and vertices can be easily worked out from the gauge fixed
action (\ref{action}), with two points worth stressing: 
first, the quadratic part (yielding the propagators) is diagonal in root 
space; second, contrary to the $SU(2)$ case, there are now vertices with 
three massive particles (corresponding to the three different roots). The 
second point is particularly crucial because it is from a diagram 
containing those vertices that we find the supergravity term.

We find twenty real massless bosons and
thirty massive complex bosons. 
We only need consider some of the latter
to construct the diagram. Writing $x_k = x_k^a H^a + x_k^\alpha E_\alpha$, with
$x_k^{-\alpha} = x_k^{\alpha *}$, we define the propagators as
\be
  \langle x_k^{\alpha *}(t_1)x_l^{\alpha}(t_2) \rangle  = 
   \Delta^{\alpha} \left( t_1, t_2 \right)_{kl} . \label{nomix}
\ee
The fluctuation $x_1$ associated to the background $R_1$  
mixes with the field $a_0$ (the fluctuation of the gauge potential). 

We focus on the vertex contained in the term of the effective 
action~(\ref{action}) of type
\be
   -2\:  \Tr \Big( [R_l, x_j][x_l, x_j] \Big) \, ,
\ee
which gives a vertex with two massive bosons and a massless one and another
one with all three massive bosons. Focusing on the second case and choosing a
particular combination of the roots we obtain a term of the type
\be
    {\bf R}^{\alpha^1} \cdot {\bf x}^{\alpha^2}\: {\bf x}^{\alpha^1}
\cdot {\bf x}^{\alpha^3}
    \equiv v^{23}\:t\: x^{13}_1x_j^{23}x_j^{12} 
    + b^{23}_l x^{13}_lx_j^{23}x_j^{12} 
     \, , \label{verti}
\ee
\FIGURE{              
\epsfxsize=6cm
\centerline{\epsfbox{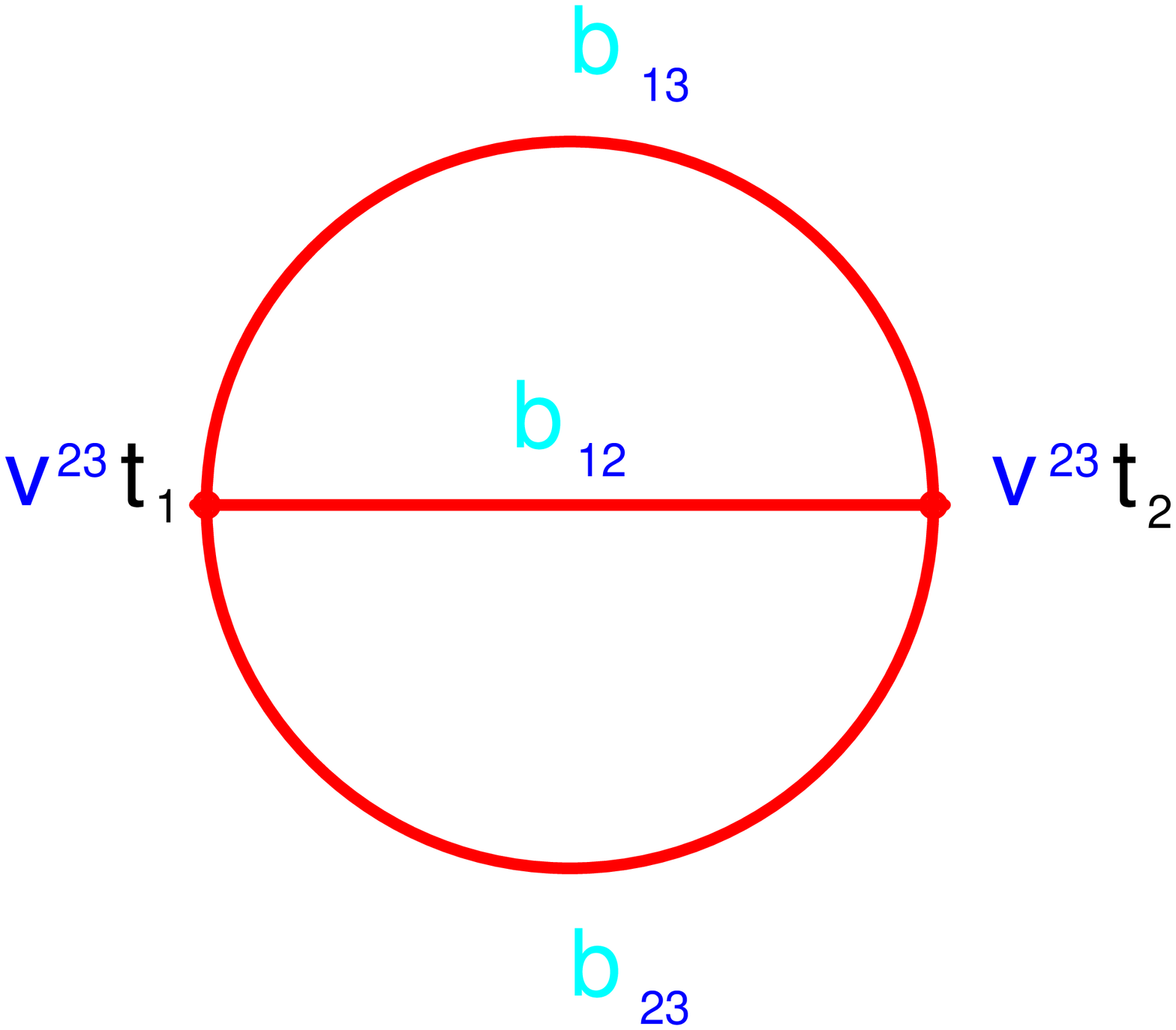}}
\caption{The setting-sun diagram.}}
The two-loop setting-sun diagram in Fig.~1, which is
obtained from the insertion of two of the vertices (\ref{verti}), yields a
contribution given, up to overall numerical factors, by
\be
a_\ominus = 
\int \di T
\int \di t\: \langle {\bf R}^{\alpha^1} (t_1)
\cdot {\bf x}^{\alpha^2 *} (t_1) \:
{\bf R}^{\alpha^1} (t_2) \cdot {\bf x}^{\alpha^2} (t_2) \rangle
\langle {\bf x}^{\alpha^1 *}(t_1) \cdot
{\bf x}^{\alpha^3 *}(t_1)
{\bf x}^{\alpha^1}(t_2) \cdot
{\bf x}^{\alpha^3}(t_2) \rangle \label{settingsun}
\ee
In eq.~(\ref{settingsun}), we defined $t=(t_1-t_2)/2$, $T=(t_1+t_2)/2$.

Our strategy 
is to do first the integration over the time difference $t$. 
Because distances are large, this leaves us with a local expression $\cal L$ 
in the ${\bf R}^{\alpha} (T)$ and their derivatives.

The propagator (\ref{nomix}) can be written as
\be
 (\Delta^{\alpha})_{kl}=
\frac{1}{R^{\alpha}} \:
\int^{(R^{\alpha})^2/\dot R^{\alpha}}_0 \frac{\di \, u \:
 e^{-u}}{\sqrt{4 \pi u}} \: 
W_{kl}(u;\: \dot R^{\alpha},\,
R^{\alpha})\:e^{-\frac{(R^{\alpha})^2}{u}t^2} \, ,\label{proptwo}
\ee 
where $W_{kl}$ is defined in the appendix B.

We note first that the integration over $u$ can be extended to
infinity, up to  exponentially negligible corrections for
$(R^{\alpha})^2/\dot R^{\alpha} \to \infty$.

The important point is that for the heavy propagators
$\Delta^{\alpha^{2}}$ and  $\Delta^{\alpha^{3}}$, we can put 
$W_{kl}=\delta_{kl}$, up to terms of order not less than
$(\dot R^{\alpha^{2,3}})^2/(R^{\alpha^{2,3}})^4$ which are not relevant for
our computation, as it will become clear in the following.
Thus, to the order in which we are interested, 
we can replace the heavy propagators with those obtained
as if $R^{\alpha^2}$ and   $R^{\alpha^3}$ were constant, that is
\be
 (\Delta^{\alpha^{2}})_{kl}= \delta_{kl}\:
\frac{1}{2 \, R^{\alpha^{2}}} e^{-2 \,R^{\alpha^{2}}|t|}\, , \label{propconst}
\ee
and similarly for  $\Delta^{\alpha^{3}}$.
Further, with the same accuracy, 
we can let $R^{\alpha^1}(t_{1}) \sim R^{\alpha^1}(t_{2})
\sim R^{\alpha^1}(T)$, since $t\leq 1/(R^{\alpha^2}+R^{\alpha^3})$, and,
for the same reason we can put 
$\Delta^{\alpha^1}(T,t)\sim\Delta^{\alpha^1}(T,t=0)$.

We can now perform the integration over $t$, obtaining 
\be
{\cal L} \sim  \frac{(R^{\alpha^1})^2}{R^{\alpha^2}R^{\alpha^3}}
\frac{1}{R^{\alpha^2} +R^{\alpha^3}}
\langle {\bf x}^{\alpha^1 *} (T) \cdot {\bf x}^{\alpha^1} (T) \rangle
  .\label{from}
\ee
where $R^{\alpha}$ still depend on $T$.

A comment is in order. The effective lagrangian~(\ref{from})
looks like a one-loop correction to the velocity
independent local effective potential:
\be 
\delta V_{1-loop}
=F\left[ R^{\alpha}(T) \right] \:
{\bf x}^{\alpha^1 *} (T) \cdot {\bf x}^{\alpha^1} (T) \, . \label{H}
\ee

One could ask whether the potential~(\ref{H})  
should vanish by supersymmetry reasons,
once one performs the sum over all the one-loop diagrams made of 
heavy fields (including the gauge, ghost and fermion fields).
The answer is: no, it is not canceled. In order to verify that this is not the
case, we have evaluated the velocity independent part of the sum of all
the diagrams with $x^{\alpha^1*}x^{\alpha^1}$ external lines, for
a generic constant background, and found that the term of eq.~(\ref{from}),
coming from the heavy loop of our setting-sun diagram, does not cancel.
In fact, the other heavy loop diagrams carry a different dependence on
$R^\alpha$.
This computation is easily done in configuration space,
because of the simple expression of the velocity independent propagators,
see eq.~(\ref{propconst}).

As a further check, 
we have also evaluated the velocity-independent effective 
potential at two loops as a function of the background---that is 
after performing the integration over the heavy
and light fields as well---and found it to be zero, as expected by
supersymmetry. Evidently, the cancellations due to the symmetry of
the theory do not occur at the level of the first step of
integrating only over the heavy fields.

Coming back to our computation,
the last propagator $\Delta^{\alpha^1}$ in~(\ref{from})
can be evaluated by expanding $W$ in 
powers of $u$ in eq.~(\ref{proptwo}).
The relevant terms are those of the kind (recall the 
overall factor $1/R^{\alpha^1}$ in front of (\ref{proptwo})): 
\be
\frac{(\dot R^{\alpha^1})^6}{(R^{\alpha^1})^{13}} \: u^6 \, .
\ee
Performing the last integral over $u$ and
substituting into (\ref{from}) we obtain 
\be
{\cal A} \equiv \int \di T \: {\cal L} = 
\int \di T \: \frac{1}{R^{\alpha^2}R^{\alpha^3} (R^{\alpha^2}+  
R^{\alpha^3})} \:
\frac{(\dot{R}^{\alpha^1})^6}{(R^{\alpha^1})^{11}} \, . \label{action2}
\ee
Notice that ${\cal A}$ is
proportional to the sixth power of $\dot R$ and therefore
is the first term in the expansion in velocities for which
no obvious supersymmetry-induced 
cancellation is expected~\cite{D2,PSS}.

To verify that the action (\ref{action2}) contains (\ref{sugra}) one 
selects terms proportional to 
$(\dot R^{\alpha^2}\dot R^{\alpha^3})^2 = v_{13}^2 v_{12}^2$,  
and performs the integration over $T$ to
get a term of the form
\be
\tilde{\cal A}\sim \frac{v_{13}^2 v_{12}^2}{B^7}\int \di \, T \: T^4 
\frac{(v_{23})^6}{\left( v^2_{23} T^2 + b^2 \right)^{11/2}} 
\sim \frac{v_{13}^2 v_{12}^2 v_{23}}{B^7b^6} \, .
\ee

The appropriate powers of $N_i$ can be
deduced---following~\cite{BBPT}--- from the double-line notation in which the
setting-sun diagram is of order $N^3$; this factor must be $N_1 N_2 N_3$
for the diagram to involve all three particles.
We thus find the term
\be
\tilde{a}_\ominus \sim
\frac{N_1 N_2 N_3 |v_{23}| v_{12}^2 v_{13}^2}{B^7b^6}
\ee
which reproduces the behavior of the
supergravity result (\ref{sugra}), that is,
$\tilde a_\ominus \sim \tilde a_\vee$.

Of course, to verify that matrix theory really matches supergravity,
one should also check the numerical coefficient as well as the matching of 
other terms with different powers of $B$, $b$ and $v_{ij}$.
That would require considering all the various graphs of the YM theory.

\acknowledgments{We thank K.\ Roland for help on the
superstring computation, E.\ Gava and K.S.\ Narain for discussions and
M.\ Dine and A.\ Rajaraman for e-mail exchanges.

Work partially supported by the Human Capital and Mobility EC program,
under contract no. ERBMRXCT 96-0045.}

\appendix

\section{The Y diagram}

The Y diagram gives a term in the amplitude of the from
\bea
A_Y & = & - 2 \: \kappa^4 \: \varepsilon_1 \cdot \varepsilon_1' \: 
\varepsilon_2 \cdot \varepsilon_2' \: \varepsilon_3 \cdot \varepsilon_3' 
\: \frac{1}{{\bf q}_1^2\: {\bf q}_2^2\: {\bf q}_3^2} \label{y} \\
& &  \times \: \Biggl\{  
({\bf p}_2 - {\bf p}_3)^2 
\Bigl[
{\bf q}_3 \cdot ({\bf p}_3 -{\bf p}_1) + 
{\bf q}_2 \cdot ({\bf p}_1 -{\bf p}_2) 
\Bigr] \Biggr. \nn \\ 
&& \quad  + \: ({\bf p}_3 - {\bf p}_1)^2
\Bigl[
{\bf q}_3 \cdot ({\bf p}_2 - {\bf p}_3) +
{\bf q}_1 \cdot ({\bf p}_1 - {\bf p}_2) 
\Bigr]  \nn \\ 
&& \Biggl. \quad  +\:  ({\bf p}_1 - {\bf p}_2)^2
\Bigl[
{\bf q}_2 \cdot ({\bf p}_2 -{\bf p}_3) +
{\bf q}_1 \cdot ({\bf p}_3 -{\bf p}_1)
\Bigr]
\Biggr\}^2 \, .  \nn
\eea
Notice that $A_Y=0$ whenever ${\bf p}_i = {\bf p}_j$, as it is also true
for $A_\vee$.
 
In the limiting 
regime~(\ref{regime}), and in our particular kinematic configuration,
the term (\ref{y}) above
turns out to give a contribution proportional to
\be
\tilde a_Y = \int \di t \: \prod_i^3 \di {\bf q}_i\; e^{i {\bf q}_i \cdot 
( \hat {\bf n}_1 v_i t 
+ {\bf b}_i)} \delta \Bigl( \sum_j {\bf q}_j \Bigr) A_Y \; =\;
f(v)\: \frac{1}{B^9 b^4} \, ,
\ee
where $f(v)$ is a homogeneous function of degree five in the velocities
$v_{ij}$, and it is therefore sub-leading with respect to $\tilde a_\vee$ 
in~(\ref{sugra}).

\section{The propagators}

The explicit form of the propagator is~\cite{BB}
\bea
\Delta^{\alpha}_{kl} & = &\int \di s \: e^{-|b^{\alpha}|^2 s} 
\sqrt{\frac{v^{\alpha}}{2\pi\sinh 2\, v^{\alpha} s}}\exp\left\{
- v^{\alpha} (t^2 \coth v^{\alpha} s
+ T^2 \tanh v^{\alpha} s) \right\} \nn \\
&& \times \left( \delta_{kl} + v^{\alpha}_kv^{\alpha}_l\:
\frac{2\, \sinh^2 v^{\alpha}s}{(v^{\alpha})^2} \right) \, . \label{prop}
\eea

By changing the integration variable to
\be
u \equiv \frac{(R^{\alpha})^2}{\dot R^{\alpha}}
\frac{\sinh v^{\alpha} s }{\cosh v^{\alpha}s},
\ee
we can rewrite the propagator~(\ref{prop}) as in eq.(\ref{proptwo}), 
with
\bea
W_{kl} & = &
\frac{1}{\sqrt{1-\frac{(\dot R^{\alpha})^2}{(R^{\alpha})^4}u^2}}
\exp \left[ -\frac{(\dot R^{\alpha})^2 (R^{\alpha})^2-
(\dot{\bf R}^{\alpha} \cdot {\bf R}^{\alpha})^2}{(R^{\alpha})^6} u^3
\sum_{n=0}
\left( \frac{\dot R^{\alpha}}{(R^{\alpha})^2} \right)^{2n}
\!\!\!\!\! \frac{u^{2n}}{3+2n} \right] \nn \\
& & \times 
\Bigg(\delta_{kl} +
\frac{\dot R^{\alpha}_k\dot R^{\alpha}_l}{(R^{\alpha})^4}
\frac{2u^2}{1-\frac{(\dot R^{\alpha})^2}{(R^{\alpha})^4}u^2}\Bigg)
\, .
\eea
The important point about $W$ is that it can be expanded in powers of
$(\dot R^{\alpha})^2/(R^{\alpha})^4$.

\clearpage
\renewcommand{\baselinestretch}{1}

\end{document}